\documentclass[twocolumn,showpacs,preprintnumbers,amsmath,amssymb,floatfix]{revtex4}
\usepackage{epsf}
\usepackage{graphicx}
 \newcommand{\insertplot}[5]{\begin{figure}
 \hfill\hbox to 0.05in{\vbox to #5in{\vfill
 \inputplot{#1}{#4}{#5}}\hfill}
 \hfill\vspace{-.1in}
 \caption{#2}\label{#3}
 \end{figure}}
 \newcommand{\inputplot}[3]{
 \special{ps: plotfile #1}
}
\begin{document}
\title{Stationary Black Holes with Static and Counterrotating Horizons}
\author{
{\bf Burkhard Kleihaus}
}
\affiliation{
{Department of Mathematical Physics, University College, Dublin,
Belfield, Dublin 4, Ireland}
}
\author{
{\bf Jutta Kunz}
}
\affiliation{
{Institut f\"ur Physik, Universit\"at Oldenburg, 
D-26111 Oldenburg, Germany}
}
\author{
{\bf Francisco Navarro-L\'erida}
}
\affiliation{
{Dept.~de F\'{\i}sica At\'omica, Molecular y Nuclear, Ciencias F\'{\i}sicas\\
Universidad Complutense de Madrid, E-28040 Madrid, Spain}
}
\date{\today}

\begin{abstract}
We show that rotating dyonic black holes with static and counterrotating
horizon exist in Einstein-Maxwell-dilaton theory
when the dilaton coupling constant exceeds the Kaluza-Klein value.
The black holes with static horizon bifurcate from the static black holes.
Their mass decreases with increasing angular momentum, their horizons are 
prolate.
\end{abstract}

\maketitle

{\sl Introduction}

Einstein-Maxwell (EM) theory admits a unique family of 
stationary axisymmetric black holes, 
the Kerr-Newman (KN) family of black holes,
characterized by their global charges:
their mass $M$, their angular momentum $J$,
their electric charge $Q$, and their magnetic charge $P$ \cite{hair}.
The central singularity is hidden behind an event horizon,
when these charges satisfy the condition
$M^2 \ge Q^2 + P^2 + J^2/M^2$.
The equality holds for extremal black holes,
which possess a finite horizon area
and vanishing surface gravity.

In many unified theories, including string theory
and Kaluza-Klein (KK) theory, dilatons appear.
Here we consider Einstein-Maxwell-dilaton (EMD) theory with action
\begin{equation}
S=\int d^4 x \sqrt{-g} \left\{ R -
{2}\partial_\mu \Phi \partial^\mu \Phi
 -  e^{2 \gamma \Phi } {\cal F}_{\mu\nu} {\cal F}^{\mu\nu} \right\}
\ . \label{action} \end{equation}
The dilaton coupling constant $\gamma$ is treated as a free parameter.
For $\gamma=0$ the dilaton decouples,
for $\gamma=1$ contact with
(the low energy effective action of) string theory is made,
for $\gamma=\sqrt{3}$ the action corresponds to KK theory \cite{emd}.

The coupling of the dilaton field to EM theory
leads to profound consequences for the black holes.
Although uncharged EMD black holes
simply correspond to EM black holes,
charged EMD black holes possess qualitatively new features.

Static EMD black holes
with only electric charge (or only magnetic charge), for instance, 
exist for arbitrarily small horizon area
\cite{emd}.
Their surface gravity 
depends in an essential way on the dilaton coupling constant $\gamma$.
In the extremal limit,
the surface gravity vanishes when $\gamma<1$,
it reaches a finite limiting value when $\gamma=1$,
and it diverges when $\gamma>1$ \cite{emd}.

Static dyonic extremal EMD black holes
can be unstable with respect to fission 
into two black holes beyond $\gamma=1$ \cite{kallosh}.
For fission to occur, 
it must be energe\-tically favourable and thermodynamically allowed.
Static EMD black holes with $|P|=|Q|$, for instance,
have a trivial dilaton field since they
correspond to Reissner-Nordstr\o m (RN) black holes,
whereas their fission products with $|P| \ne |Q|$
possess a non-trivial attractive scalar field, 
lowering their masses.

Rotating charged EMD black holes are known exactly only for
KK theory \cite{FZB,Rasheed}.
For arbitrary dilaton coupling constant $\gamma$,
only perturbative results are available,
for only electrically charged black holes \cite{HH,casadio}.

Dyonic extremal KK black holes possess further
surprising properties \cite{Rasheed}.
When $(P/M)^\frac{2}{3}+(Q/M)^\frac{2}{3}=2^\frac{2}{3}$,
they can carry finite angular momentum $|J|\le |PQ|$, 
while they possess a vanishing horizon angular velocity $\Omega$
and no ergoregion. For these black holes
the angular momentum can be increased from zero to $|PQ|$
without changing their mass or their charges.

Here we show that for dilaton coupling $\gamma> \sqrt{3}$,
whole families of non-extremal black holes
with finite angular momentum $|J|\le |PQ|$ exist,
which possess vanishing horizon angular velocity and no ergoregion.
Moreover, we demonstrate, that even stranger black holes
appear: black holes with counterrotating horizon,
never encountered before.

{\sl Ansatz and Boundary Conditions}

We consider stationary, axially symmetric black hole space times
with Killing vector fields, $\xi=\partial_t$, $\eta=\partial_{\varphi}$.
The Lewis-Papapetrou form of the metric reads in isotropic coordinates
\begin{equation}
ds^2 = -fdt^2+\frac{m}{f}\left[dx^2+x^2 d\theta^2\right] 
       +\sin^2\theta x^2 \frac{l}{f}
          \left[d\varphi-\frac{\omega}{x}dt\right]^2 \  
 . \label{ansatzM} \end{equation}
The gauge field is parametrized by
\begin{eqnarray}
{\cal A}_\mu dx^\mu &=& 
 {\cal A}_0 dt + {\cal A}_{\varphi} d\varphi \ . \label{ansatzA}
\end{eqnarray}
The functions $f$, $m$, $\omega$,
$\Phi$, ${\cal A}_0$, and ${\cal A}_{\varphi}$ 
depend on $x$ and $\theta$, $l$ depends on $x$ alone.

The event horizon 
resides at a surface of constant radial coordinate, $x=x_{\rm H}$,
and is characterized by the condition $f(x_{\rm H})=0$ \cite{kkrot}.
The Killing vector field $\chi = \xi + {\Omega} \eta$
is orthogonal to and null on the horizon \cite{wald}. 
We note, that the equations scale with $x_{\rm H}$.

At the horizon, we impose the boundary conditions
$f=m=l=0$, 
$\omega=\Omega x_{\rm H}$,
$\partial_x \Phi=0$,
${\cal A}_0+ \Omega {\cal A}_{\varphi}=\Psi_{\rm e,H}$, and
$\partial_x {\cal A}_{\varphi}=0$.
Here $\Psi_{\rm e,H}$ denotes the horizon electrostatic potential.

At infinity we impose the boundary conditions
$f=m=l=1$, $\omega=0$,
$\Phi=0$,
${\cal A}_0=0$,
${\cal A}_{\varphi}= P \cos\theta$.

On the symmetry axis, we impose
for $\theta=0$ the boundary conditions
$\partial_\theta f = \partial_\theta m = \partial_\theta l =
\partial_\theta \omega = 0$,
$\partial_\theta \Phi =0$,
$\partial_\theta {\cal A}_0 =0$,
${\cal A}_{\varphi} = P$.
The boundary conditions for $\theta=\pi$ 
coincide with these, except for 
${\cal A}_{\varphi} = - P$.

Regularity on the symmetry axis requires $m=l$.
The function $l$ can be solved for analytically.
For rotating dyonic black holes
the symmetry with respect to reflection at the equatorial
plane, present for black holes with only electric charge
(or only magnetic charge), is broken.

{\sl Black Hole Properties}

The global charges of the black holes
can be read off the asymptotic expansion \cite{emdlong},
$$
 f \rightarrow 1-\frac{2M}{x} \ , \ \ \ \omega \rightarrow \frac{2 J}{x^2}
 \ , \ \ \
 \Phi \rightarrow -\frac{D}{x} \ ,
$$
\begin{equation}
{\cal A}_0 \rightarrow \frac{Q}{x} \ , \ \ \ 
{\cal A}_\varphi \rightarrow P \cos \theta 
\ . \end{equation}
The dilaton charge $D$ depends on the electromagnetic charges, and vanishes
when $|P|=|Q|$ 
\cite{emd,FZB,Rasheed}.

The area parameter $x_\Delta$
is defined via the black hole horizon area $A=4 \pi x_\Delta^2$.
The surface gravity $\kappa$ is obtained from 
the Killing vector $\chi$ \cite{wald}
\begin{equation}
\kappa^2 = - \frac{1}{4} (\nabla_\mu \chi_\nu)(\nabla^\mu \chi^\nu)
\ . \label{sgwald} \end{equation}
$\kappa$ is constant at the horizon,
as required by the zeroth law of black hole mechanics.
A measure for the deformation of the horizon is given by
the ratio of the circumference of the horizon along the equator, 
${L}_{\rm e}$, and the circumference of the horizon along the poles,
${L}_{\rm p}$.

EMD black holes satisfy the Smarr formula \cite{smarr,Rasheed,KKN},
as well as the mass formula \cite{KKN}
\begin{equation}
M =  \frac{\kappa A}{4\pi}
 + 2 \Omega J + 2\Psi_{\rm e,H} Q + \frac{D}{\gamma}
\ , \label{namass} \end{equation}
valid even for non-Abelian black holes.

{\sl Results}

Here we investigate rotating EMD black holes numerically
\cite{fidisol}.
We first consider black holes with static horizon,
i.e.~$\Omega=0$, but non-vanishing angular momentum $J$.
These black holes are necessarily dyonic.
They bifurcate from the static spherically symmetric black holes
at critical values of the parameters.

In Fig.~1 we show the angular momentum $J$ of such rotating 
$\Omega=0$ black holes,
bifurcating from the static black holes,
for fixed charges $P=Q$ and several values of the horizon radius $x_{\rm H}$
as a function of the dilaton coupling constant $\gamma$.
The figure demonstrates that the upper limit of the angular momentum,
$|J|=|PQ|$, valid for extremal $\Omega=0$ KK black holes, 
remains an upper limit
also for non-extremal $\Omega=0$ black holes (where $\gamma>\sqrt{3}$).
For horizon radius $x_{\rm H}\rightarrow 0$,
$J$ becomes increasingly steep and tends to a vertical line 
in the KK limit.
There are no rotating $\Omega=0$ black holes for $\gamma<\sqrt{3}$.
\begin{figure}[h!]
\begin{center}
\epsfysize=6.5cm
\mbox{\epsffile{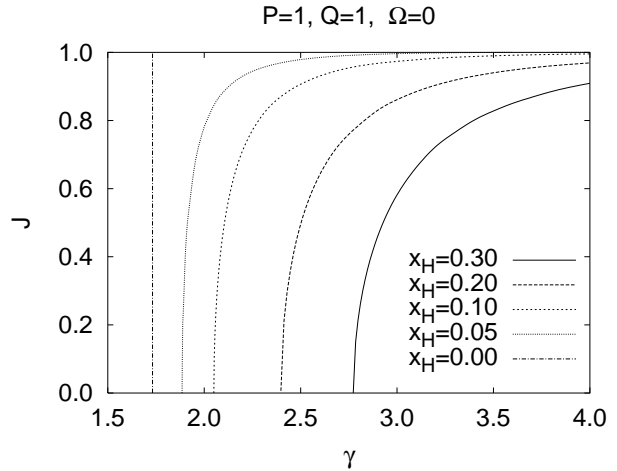}}
\caption{ 
Angular momentum $J$ versus dilaton coupling $\gamma$
for rotating $\Omega=0$ black holes
($P=Q=1$, $x_{\rm H}=0$, 0.05, 0.1, 0.2, 0.3).
}
\end{center}
\end{figure}

For $|P|=|Q|$ the static black holes correspond to RN black holes.
Therefore the critical values of the parameters, where the rotating $\Omega=0$
black holes bifurcate from the static black holes, can readily be
obtained perturbatively.
The charges scale with $x_{\rm H}$,
and consequently also the critical values of the parameters.
${\cal P}_{\rm cr}=(P/x_{\rm H})_{\rm cr}$, for instance,
yields the critical charge $P_{\rm cr}$ for a given horizon, 
or the critical horizon $x_{\rm H, cr}$ for a given charge.
The dependence of ${\cal P}_{\rm cr}$ on $\gamma$ is shown in Fig.~2.
Again we see, that $\gamma = \sqrt{3}$
represents the limit for rotating black holes with static horizon.
\begin{figure}[h!]
\begin{center}
\epsfysize=6.5cm
\mbox{\epsffile{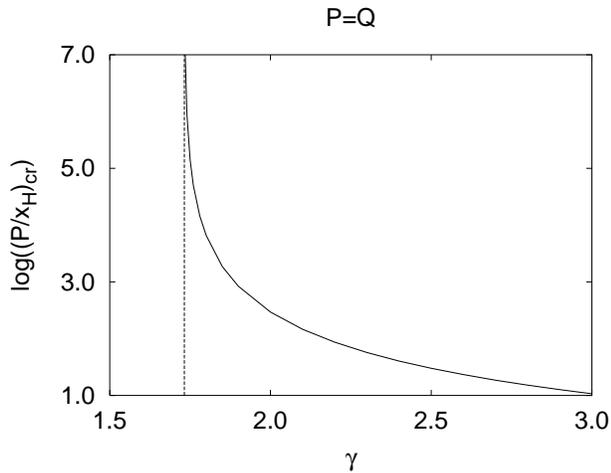}}
\caption{ 
$\gamma$-dependence of the
critical values of the scaled charge 
${\cal P}_{\rm cr}=(P/x_{\rm H})_{\rm cr}$
for rotating $\Omega=0$ black holes
($P=Q$). The vertical line indicates $\gamma=\sqrt{3}$.
}
\end{center}
\end{figure}

Considering rotating $\Omega=0$ black holes 
with fixed $\gamma$, $P$, and $Q$,
and varying horizon radius $x_{\rm H}$, we observe a monotonic
decrease of the angular momentum $J$ from its extremal value
$|J|=|PQ|$ at $x_{\rm H}=0$ to zero at the critical value $x_{\rm H, cr}$,
the bifurcation point with the static black holes.
The horizon angular momentum $J_{\rm H}$ is always negative, 
and vanishes only at the end points.
At the same time,
the horizon area and the mass increase monotonically
from their extremal values at $x_{\rm H}=0$
to the static black hole values at $x_{\rm H, cr}$.
The horizon area of the extremal solution at $x_{\rm H}=0$ vanishes.
Interestingly, the mass of the rotating $\Omega=0$ black holes
is always smaller than the mass of the
static black holes with the same horizon area (when these exist).
Moreover the horizon of these black holes has a prolate deformation,
whereas the horizon of KN black holes is always oblate. 

The presence of non-extremal rotating $\Omega=0$ black holes 
immediately raises the question, what properties 
neighbouring black holes with finite $\Omega$ possess.
Intriguingly, we here encounter 
black holes with counterrotating horizon, 
i.e.~the horizon angular velocity $\Omega$
and the global angular momentum $J$ have opposite signs.
Thus the rotating $\Omega=0$ black holes correspond to boundaries
in parameter space, beyond which black holes with counterrotating
horizon reside.

This is illustrated in Fig.~3, where we exhibit
the angular momentum $J$ as a function of the electric charge $Q$
for rotating black holes with fixed magnetic charge,
horizon radius, and dilaton coupling 
for several values of the horizon angular velocity $\Omega$.
The rotating $\Omega=0$ black holes bifurcate 
from the static ($J=0$) black holes
at the critical value $Q_{\rm cr}$.
Black holes with corotating horizon, where $J$ and $\Omega$
have the same sign, reside above the $\Omega=0$ curve, 
black holes with counterrotating horizon reside between the $\Omega=0$ curve
and the static curve.
\begin{figure}[h!]
\begin{center}
\epsfysize=6.5cm
\mbox{\epsffile{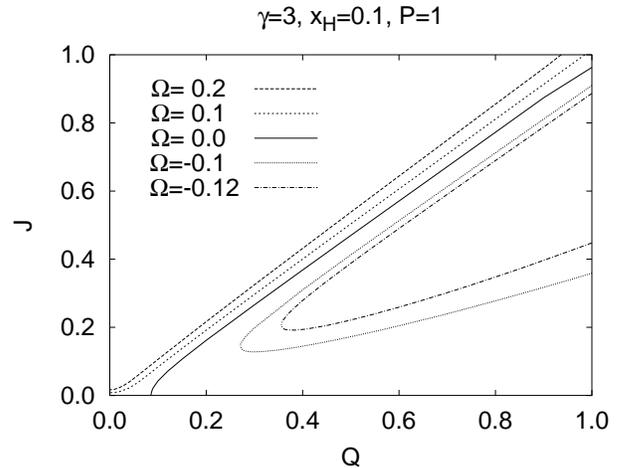}}
\caption{ 
Angular momentum $J$ versus electric charge $Q$
for black holes with fixed horizon angular velocity
$\Omega=0.2$, 0.1, 0, -0.1, -0.12
($P=1$, $x_{\rm H}=0.1$, $\gamma=3$;
$Q_{\rm cr}=0.085$)
}
\end{center}
\end{figure}

To gain a better understanding of these strange counterrotating black holes,
let us consider their global and horizon properties.
We exhibit in Fig.~4 the black hole mass $M$ and angular momentum $J$ 
as functions of the horizon angular velocity $\Omega$
starting from the static black hole,
for several values of the horizon radius $x_{\rm H}$,
and fixed $\gamma$, $P$, and $Q$.
Counterrotating black holes then exist, when $x_{\rm H}<x_{\rm H, cr}$.
(Note, that an energetically degenerate set of black holes is obtained
for $\Omega \rightarrow -\Omega$, $J \rightarrow -J$.)
\begin{figure}[h!]
\begin{center}
\epsfysize=6.5cm
\mbox{\epsffile{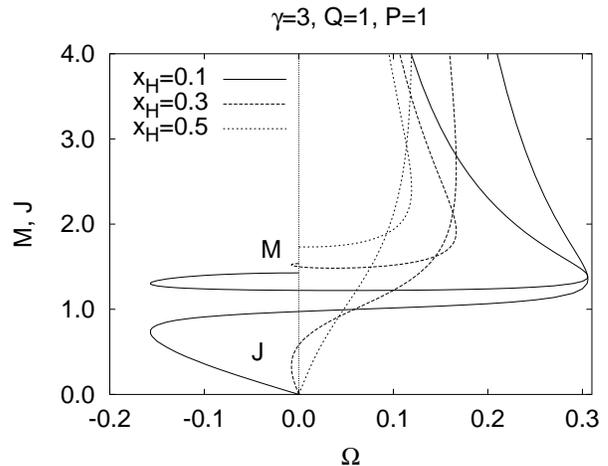}}
\caption{ 
Black hole
angular momentum $J$ and mass $M$ versus horizon angular velocity $\Omega$
($P=Q=1$, $x_{\rm H}=0.1$, 0.3, 0.5, $\gamma=3$;
$x_{\rm H, cr}=0.357$).
}
\end{center}
\end{figure}

As seen in Fig.~4,
when counterrotating EMD black holes emerge from the static black hole,
the horizon velocity $\Omega$ first decreases with increasing $J$,
reaches a minimum at $\Omega_{\rm min}$, and then increases, 
passing a rotating $\Omega=0$ black hole.
The horizon velocity reaches a maximum at
$\Omega_{\rm max}$, and then decreases again, 
tending finally to zero.
Along this path, the angular momentum $J$ increases monotonically,
from zero to infinity.

The mass $M$, in contrast, decreases from the mass of the static black hole,
and continues to decrease
monotonically even beyond the rotating $\Omega=0$ black hole,
up to a point $\Omega^*>0$. 
Thus the mass decreases with increasing $J$ 
not only for counterrotating black holes but even for the adjacent
black holes with corotating horizon.
Beyond $\Omega^*$ the mass increases monotonically.

Noting that $\kappa A= 8\pi x_{\rm H}$, for fixed $x_{\rm H}$
a lowering of the mass is expected 
from the mass formula (\ref{namass})
for counterrotating black holes,
since $\Omega J<0$.
It is surprising, however, that the mass continues to decrease 
beyond $\Omega=0$, since $\Omega J>0$. 
Here the decrease of the mass is
due to a more than compensating decrease of $\Psi_{\rm e, H} Q$.
In fact, $\Psi_{\rm e, H}$ decreases strongly, and almost linearly between
$\Omega_{\rm min}$ and $\Omega_{\rm max}$. ($D=0$, when $|P|=|Q|$.)

In Fig.~5 we exhibit horizon properties
for the black holes of Fig.~4.
For counterrotating black holes
the shape of the black hole horizon
is oblate close to the static black hole,
but turns prolate in the vicinity of $\Omega_{\rm min}$,
and remains prolate approximately up to $\Omega_{\rm max}$,
reaching the strongest prolate deformation at $\Omega\approx \Omega^*$. 

Considering fission of black holes, we conclude,
that static dyonic extremal black holes may not only be unstable
with respect to fission into two static extremal black holes
\cite{kallosh},
but that they may be unstable as well
with respect to fission into two extremal rotating black holes \cite{Rasheed},
also for $\gamma > \sqrt{3}$.

 \begin{figure}[h!]
\begin{center}
\epsfysize=6.5cm
\mbox{\epsffile{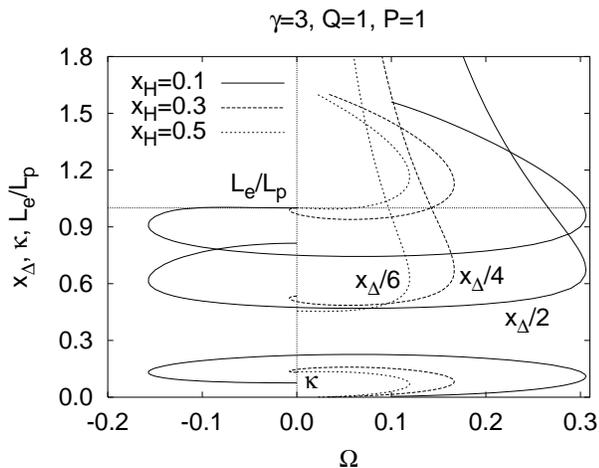}}
\caption{  
Same as Fig.~4 for
the horizon properties $x_\Delta$, $\kappa$, and $L_e/L_p$.
The horizontal line indicates $L_e/L_p=1$.
}
\end{center}
\end{figure}

Let us give a summary of our results, 
restricting to black holes with $|P|=|Q|$ in Fig.~6.
Here the scaled angular momentum $|J|/M^2$ of the extremal EMD black holes is shown
versus the scaled charge $|P|/M$
for several values of $\gamma$ \cite{Rasheed,foot}.
For a given $\gamma$ black holes only exist in the regions bounded by the 
axes and the solid curves. 
The dotted curves correspond to the rotating $\Omega=0$ black holes.
For $\gamma>\sqrt{3}$ the corotating black holes are to the left 
of the dotted curve, whereas the counterrotating black holes are in 
the shaded region to the right of the dotted curve. As $\gamma$ approaches
$\sqrt{3}$ the shaded region degenerates to the line 
forming the vertical part of the boundary, where only rotating $\Omega=0$
solutions exist. For $\gamma<\sqrt{3}$ no rotating $\Omega=0$
or counterrotating black holes exist.
The dashed curve indicates the location of the spike of the boundary for
$\gamma>\sqrt{3}$
\begin{figure}[h!]
\begin{center}
\epsfysize=6.5cm
\mbox{\epsffile{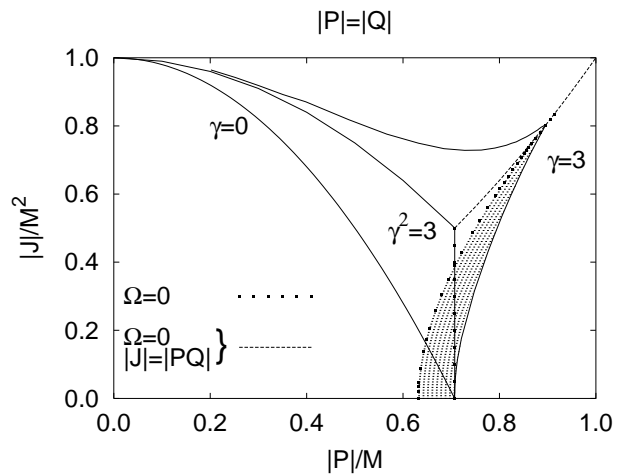}}
\caption{ 
Scaled angular momentum $J/M^2$ versus 
scaled charge $P/M$ for extremal black holes
and rotating $\Omega=0$ black holes
($P=Q$, $\gamma=0$, $\sqrt{3}$, 3).
}
\end{center}
\end{figure}

Finally, we conjecture that black holes with counterrotating horizon
may exist also in other theories.


\begin{thebibliography}{000}
\bibitem{foot}
 For $\gamma > \sqrt{3}$, the extremal black holes are
 approximated by almost extremal black holes for numerical reasons.
\end{thebibliography}
\end{document}